\documentclass[aps,prd,preprint,fleqn,showkeys,showpacs,superscriptaddress]{revtex4}
\usepackage[colorlinks=true]{hyperref}
\hypersetup{colorlinks=true, citecolor=blue, urlcolor=blue, linkcolor=blue}

\usepackage{graphicx}
\usepackage{tabulary}
\usepackage{multirow,tabularx}
\usepackage{amssymb}
\usepackage{amsmath}
\usepackage{bm}
\usepackage{enumitem}
\usepackage{etoolbox}
\usepackage{xcolor}
\usepackage{soul}

\usepackage{subcaption}

\makeatletter

\patchcmd\frontmatter@PACS@format{\addvspace{11\p@}}{}{}{}
\pretocmd\frontmatter@keys@format{\addvspace{11\p@}}{}{}
\patchcmd{\titleblock@produce}
  {\@pacs@produce\@pacs\@keywords@produce\@keywords}
  {\@keywords@produce\@keywords\@pacs@produce\@pacs}
  {}{}
\makeatother

\DeclareUnicodeCharacter{2212}{-}
\begin{document}
\title{Investigations of the semileptonic decays of $D$ and $D_s$ into a pseudoscalar meson using the Isgur-Wise Functions}

\author{S. Rahmani}
\email{s.rahmani120@gmail.com}
\affiliation{School of Physics and Electronics, Central South University, Changsha 410083, China}

\author{C. W. Xiao}
\email{xiaochw@csu.edu.cn}
\affiliation{School of Physics and Electronics, Central South University, Changsha 410083, China}

\begin{abstract} 
The branching ratios of the semileptonic decay widths of the charm mesons are analyzed, using three different models for the Isgur-Wise functions, such as ${D^0} \to {K^ - }{l^ + }\upsilon $, ${D^0} \to {\pi ^ - }{l^ + }\nu $, ${D_s} \to {K^0}{l^ + }\nu $  and ${D_s} \to \eta {l^ + }\nu$, where the form factors of these decays are discussed. The mass spectra of the charm mesons are obtained. We use a potential quark model and consider the non-relativistic Hamiltonian of the charm meson as a bound state of the quark-antiquark system. We take into account the harmonic-type confinement and also Hellmann potential, which is a superposition of the Coulomb and the Yukawa potential. Using the variational approach along with the harmonic oscillator wave functions, we evaluate the mass spectra of the charm mesons, the form factors and the semileptonic decay widths of $D_{(s)}$. We present our results for masses of $D, D_s$ and $\eta$, the Isgur-Wise functions, the form factors of the semileptonic decays, and the branching fractions of the semileptonic decays of $D$ and $D_s$. Our results are motivating.
\end{abstract}

\maketitle
\thispagestyle{empty}

\section{Introduction }

The transition form factors of the heavy mesons semileptonic decays are important in extracting the CKM matrix elements such as $|{V_{ub}}|,|{V_{cb}}|,|{V_{cd}}|$  and $|{V_{cs}}|$. For instance, the differential decay rate $\frac{{d\Gamma (D \to Kl\nu )}}{{d{q^2}}}$ is proportional to $|{V_{cs}}{|^2}$ and $F({q^2})$, where $q^2$ is the square of four momentum transfer corresponding to the mass of the virtual $W$ boson and $F(q^2)$ is the transition form factor, which is related to momentum transfer \cite{Wirbel:1985ji}. To obtain the transition form factor, the knowledge of internal structure of the hadrons is essential. $D$ meson consists of a charm quark ($c)$ and a light quark, where a heavy quark interacts with light degrees of freedom and the heavy quark effective theory \cite{Neubert:1993mb} should be considered, which is a successful theory for studying the weak decays of heavy hadrons. In 1998, Khodjamirian derived the form factors for the transition $D \to \pi $ using the light cone QCD sum rules \cite{Khodjamirian:1998vk}. In 2005, Aubin et al. had presented the semileptonic decay form factors of the $D$ to $K$ and $D$ to $\pi$ transitions with the lattice QCD \cite{FermilabLattice:2004ncd}. Based on the covariant light-front quark model, Cheng and Kang obtained the decay rates of semileptonic $D$ and $D_s$ decays \cite{Cheng:2017pcq}. The form factors of the semileptonic decays of $D$ to $K$ had been determined by Zhang et al. \cite{Zhang:2018jtm} in 2018. Ivanov et al. provided a theoretical description of the charm mesons semileptonic decays based on the covariant confining quark model and constructed the matrix element for semileptonic decays of $D_{(s)}$ in terms of different form factors \cite{Ivanov:2019nqd}. The weak decay form factors for ${D_{(s)}} \to P(V)\ell {\nu _\ell }$ had been evaluated by Faustov et al. using the quasipotential approach and the relativistic quark model \cite{Faustov:2019mqr}. Yao et al. computed semileptonic transition form factors of $D_s^{} \to K$, $D \to K$ and $D \to \pi $ with the quantum field theory \cite{Yao:2020vef}. The different observables in the semileptonic decays of the type  ${D_{(s)}} \to P(V){l^ + }{\nu _l}$ were evaluated by Zhang et al. using the covariant light-front quark model \cite{Zhang:2020dla}. Indeed, the study of semileptonic transitions of the charm mesons is still challenging and interesting in the flavor physics due to the various parameterizations of the form factors. We follow the parameterization of the form factors regarding Isgur-Wise Function (IWF) \cite{Isgur:1989vq,Isgur:1988gb,Isgur:1990yhj,Coleman:2000gu}, which is a powerful approach in analyzing of the heavy mesons semileptonic weak decays. 
\par
Due to a large number of the measurements on the charm and bottom sector, the structure of charm and bottom heavy-light mesons has been paid much theoretical attention. Thus, the spectrum and decay properties of charm mesons are well studied in the literature. Leptonic decays of heavy pseudoscalar mesons were investigated in the quark model \cite{Gershtein:1976aq}. In 1978, Khlopov examined ${P_1} \to {P_2}l\nu $ pseudoscalar meson decays \cite{Khlopov:1978id}. $D$ and $D_s$ mesons were studied with a flux tube model \cite{Shan:2008ga}, where a meson is assumed as a massive quark and one massive antiquark connected by a flux tube (or relativistic string) with universal constant tension. Bhaghyesh obtained the mass spectra of the charmonium using a modified Cornell potential with a Gaussian-smeared contact hyperfine interaction \cite{A:2021vdw}. The mass spectra of the charmed-strange mesons as well as the strong decay widths were investigated by Gao et al. \cite{Gao:2022bsb} with the modified relativized quark model. Zhao et al. calculated the masses of low-lying S-wave mesons including charm ones using the harmonic oscillator wave function along with the Cornell-like potential and one-gluon exchange spin-spin interaction \cite{Zhao:2021cdv}. In our previous work \cite{Xiao:2020gry}, we applied one gluon exchange plus screened potential and obtained the masses of heavy-light mesons. In the present work, we take a modified harmonic potential with a Gaussian-type wave function to study the mass and decay properties of the charm mesons. 
\par
The charm meson spectroscopy can be used for testing the quark model. In the present work, building a model for the meson structure, we apply a phenomenological potential model to calculate the spectrum of charm mesons. By using the obtained masses, we investigate different form factors of the semileptonic transitions of $D$ and $D_{s}$, in order to extend our phenomenological potential model to study the semileptonic decay widths of $D$ and $D_s$. Our work is organized as follows. In the next section, we start from the wave function of mesons and evaluate the charm meson mass spectrum. In section \ref{section:3},  we investigate the semileptonic decays ${D^0} \to {K^ - }{e^ + }\nu $, ${D^0} \to {K^ - }{\mu ^ + }{\nu _\mu }$, ${D^0} \to {\pi ^ - }{e^ + }{\nu _e}$ , ${D^0} \to {\pi ^ - }{\mu ^ + }{\nu _\mu }$, $D_s^{} \to {K^0}{e^ + }{\nu _e}$ , $D_s^{} \to \eta {e^ + }{\nu _e}$ and $D_s^{} \to \eta {\mu ^ + }{\nu _\mu }$  using three different IWFs. We close with a summary in Section \ref{section:4}. 
\section{Theoretical Framework}
\label{section:2}
\begin{figure}
\centering
\includegraphics[width=0.48\linewidth]{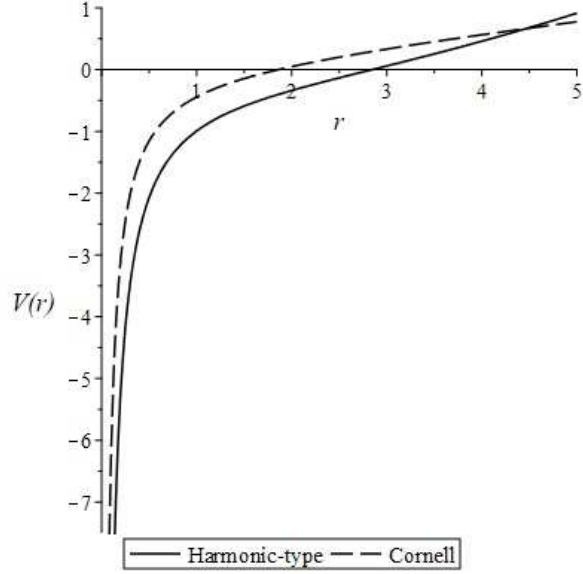}
\caption{Considered potential, Eq. (\ref{eq:potential}) in comparison with the Cornell potential $V(r) = \frac{{ - 4{\alpha _s}}}{{3r}}+br$ \cite{Barnes:2005pb}.} 
\label{fig:potential}
\end{figure}

We should choose a suitable potential that describes the internal interactions of the quarks inside the hadrons. There are several QCD potential quark models for describing the bound state of quark-antiquark, which should be taken into account two important properties of the QCD, the confinement and asymptotic freedom. There are some typical potential models, such as the screened potential \cite{Xiao:2020gry}, the Deng-Fan–type \cite{Hassanabadi:2015scl} and the Hulthen type \cite{Bhaghyesh:2011zza}. For the confinement part, one can take a linear, harmonic and exponential potential, that give a satisfactory description of the quark-antiquark bound states \cite{Lucha:1991vn, Xiao:2020gry, Bhaghyesh:2011zza, Li:2009zu, Hassanabadi:2015scl}. Among different types of phenomenological potential models, we consider a combination of the harmonic confinement, the Coulomb interaction, and the Yukawa terms. The Yukawa potential ($\beta \frac{{{e^{ - \alpha r}}}}{r}$) was originally proposed for the exchange interaction, which was also known as the screened Coulomb potential \cite{Alhaidari2008}. The superposition of the Yukawa and Coulomb terms, $\frac{{ - 4{\alpha _s}}}{{3r}} + \beta \frac{{{e^{ - \alpha r}}}}{r}$, was named as the Hellmann potential \cite{Nasser:2011} and had dealt with the cancellation of the attractive and repulsive interactions at short distances. A confinement term is also needed, which is responsible for the interactions at long distances. The linear and harmonic confining terms had been applied to study mesonic systems \cite{Lengyel:2000dk, Yang:2011ie, Rahmani:2017vbg}. Thus, by taking into account all these contributions, we take an interaction potential with the type of following,
\begin{equation}
    V(r) = \frac{{ - 4{\alpha _s}}}{{3r}} + \beta \frac{{{e^{ - \alpha r}}}}{r} + \frac{1}{2}\mu {\kappa ^2}{r^2} + {V_0},
    \label{eq:potential}
\end{equation}
where $\beta$, $\alpha$, $\kappa$ and $V_0$ are potential parameters. By fitting to the experimental mass spectrum of $D$, $D_s$ and the other charmonium, the potential parameters have been chosen as  $\beta  =  - 0.42$, $\alpha = 0.01$ GeV and $\kappa = 0.34$ GeV. We have also used the values of $V_{0}$ parameter as, ${V_0} = 0.142 $ GeV,  ${V_0} =$ 0.114 GeV,  ${V_0} = 0.628$ GeV, ${V_0} = $  0.545 GeV, ${V_0} = $ -0.161 GeV and  ${V_0} = $ -0.105 GeV associated to the states ${1^1}{S_0}$, ${1^3}{S_1}$, ${2^1}{S_0}$, ${2^3}{S_1}$, ${1^3}{P_0}$ and ${1^3}{D_3}$, respectively. $\mu$ is the reduced mass of the meson system and ${\alpha _s}$ the strong running coupling constant, given by 
\begin{equation}
    {\alpha _s} = \frac{{4\pi }}{{(11 - \frac{2}{3}{n_f})\ln (\frac{{{{(2\mu )}^2} + 1}}{{\Lambda _{QCD}^2}})}},
    \label{alpha_s}
\end{equation}
with ${\Lambda _{QCD}} = 0.413$ GeV, where $\alpha_s$ is responsible for the fundamental coupling underlying the interactions of quarks and gluons in QCD, $n_f$ the number of quark flavors. In fact, Eq. (\ref{alpha_s}) is originated from one loop expression $\alpha _B^{(1)}({Q^2}) = \frac{{4\pi }}{{{\beta _0}\ln (\frac{{{Q^2} + M_B^2}}{{{\Lambda ^2}}})}}$ with ${\beta _0} = 11 - \frac{2}{3}{n_f}$ \cite{Badalian:2001by}, where $M_B$ is the background mass, which has the value about 1 GeV. Taking $Q=(2\mu)^2$ \cite{Deur:2016tte}, we will have Eq. (\ref{alpha_s}). In Eq. (\ref{eq:potential}), the Yukawa and Coulomb terms behave as short range interactions, whereas the confinement harmonic interactions stand for the long range ones. The behavior of Eq. (\ref{eq:potential}) is depicted in Fig. \ref{fig:potential}, where the harmonic-type of the solid line is the one of Eq. (\ref{eq:potential}) and the Cornell potential of the dashed line is the superposition of the linear confinement and the Coulomb terms, $\frac{{ - 4{\alpha _s}}}{{3r}}+br$, with $b = 0.18$ $GeV^2$ \cite{Barnes:2005pb}. From Fig. \ref{fig:potential}, one can see that the interactions are stronger for large values of $r$, which are mainly contributed from the harmonic part, whereas, they become weaker at the short distance of $r$, which are dominant by the Yukawa and Coulomb part, as we have expected from the confinement and asymptotic freedom.

\begin{figure}
\centering
\includegraphics[width=0.48\linewidth]{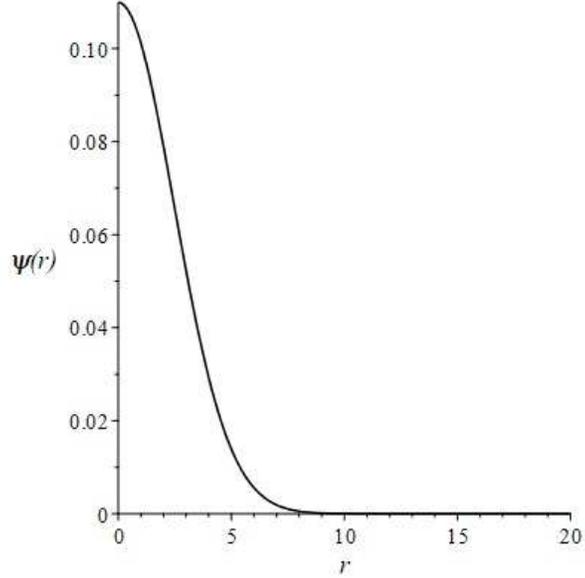}
\caption{Wave function of the $D$ meson of the ground state.}
\label{fig:wave D}
\end{figure}

\begin{figure}
\centering
\includegraphics[width=0.48\linewidth]{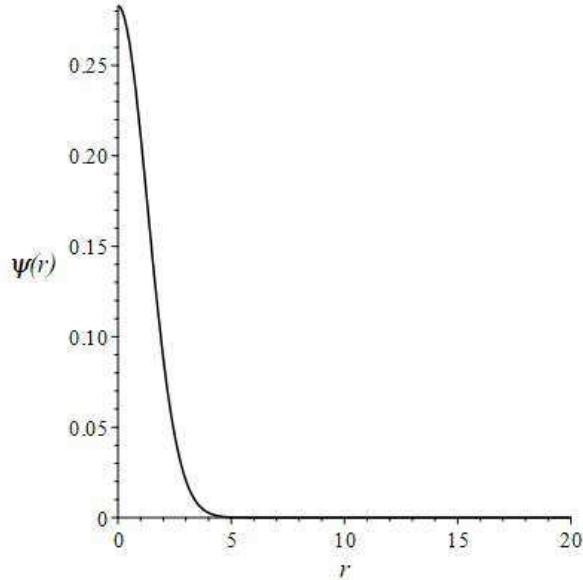}
\caption{Wave function for the $J/ \psi$ state.}
\label{fig:wave J}
\end{figure}

Using the variational approach \cite{Lucha:1991vn} and considering the confined part as a dominant part in the potential, the wave function is taken as the harmonic oscillator radial wave function, given by \cite{Pang:2017dlw}
\begin{equation}
    {\psi _{n,l}}(a,r) = N{(ar)^l}{e^{ - {a^2}{r^2}}}L_{n - 1}^{(l + \frac{1}{2})}({a^2}{r^2}),
    \label{wave function}
\end{equation}
where $n,l$ are the principal and the angular momentum quantum numbers of physical states, $a$ is the variation parameter, $N$ the normalization constant and $L_{n - 1}^{(l + \frac{1}{2})}({a^2}{r^2})$ represents the Laguerre polynomial. Note that the long-distance behaviours of the wave functions are significant and dominant, so that the harmonic oscillator wave function is legitimate \cite{Hassanabadi:2015scl}. Besides, taking the harmonic confining term in Eq. (\ref{eq:potential}) as the parent part, one can obtain mesonic wave function of Eq. (\ref{wave function}) by using a non-relativistic Hamiltonian, given by
\begin{equation}
    H =  - \frac{{{\hbar ^2}}}{{2\mu }}{\nabla ^2} + V(r).
    \label{Hamilton}
\end{equation}
 Thus, applying the variational approach, the variation parameter ($a$) is obtained by minimizing the energy of the system,
\begin{equation}
    {E_{n,l}(a)} = \frac{{\left\langle {\psi_{n,l}(a,r) |H|\psi_{n,l}(a,r) } \right\rangle }}{{\left\langle {\psi_{n,l}(a,r) |\psi_{n,l}(a,r) } \right\rangle }}.
    \label{energy}
\end{equation}
Then, the mass of the mesonic systems is taken to be
\begin{equation}
    M = {m_1} + {m_2} + {E_{n,l}} + \left\langle {{V_{sd}}} \right\rangle ,
    \label{mass}
\end{equation}
where $m_1$ and $m_2$ are the masses of quark constituents, ${E_{n,l}}$ is the energy of the system, given by Eq. (\ref{energy}), and the spin dependent interaction, ${V_{sd}}$, is added perturbatively. $\left\langle {{V_{sd}}}\right\rangle$ means the expectation value of $V_{sd}$, where the spin dependent interaction is taken as \cite{Lengyel:2000dk}
\begin{equation}
    {V_{sd}} = \frac{2}{{3{m_1}{m_2}}}({{\vec s}_1}.{{\vec s}_2})(4\pi {\alpha _s}\delta (r) + 6A),
\end{equation}
where the notation $\left\langle {{\vec s}_1}.{{\vec s}_2} \right\rangle  = \frac{1}{2}(S(S + 1) - {s_1}({s_1} + 1) - {s_2}({s_2} + 1))$ is used in the $\left\langle {{V_{sd}}} \right\rangle $ in Eq. (\ref{mass}) and $A = 0.014$ $GeV^3$, which is taken from Ref. \cite{Lengyel:2000dk}. Note that $s_1$ and $s_2$ are the spins of the constituent quarks. Thus, $S$ refers to the total spin of the mesonic system, $\vec S = {{\vec s}_1} + {{\vec s}_2}$ which has the value zero for spin singlets and one for spin triplets. For the calculation of the mass of the mesonic system having $\delta (r)$ for the spin dependent part, we should consider softening their singularity by a quasistatic approximation. Thus, we use the approximation of the type \cite{Radford:2009bs}
\begin{equation}
    \delta (r) = \frac{{{{\omega '}^2}}}{{\pi r}}{e^{ - 2\omega 'r}},
\end{equation}
with ${\omega '^2} = \frac{{2m_1^2m_2^2}}{{m_1^2 + m_2^2}}$. Besides, the input quark masses are taken as ${m_s} = {\text{0}}{\text{.483}}$ GeV, ${m_u} = {m_d} = 0.336$ GeV and ${m_c} = 1.55$ GeV \cite{Hassanabadi:2015scl}. We show our results for the masses of the charmed mesons including $D$, $D_s$ and charmonium mesons in Tables \ref{tab:massD}, \ref{tab:massDs}, \ref{tab:masseta}, respectively. As one can be seen from Tables \ref{tab:massD}, \ref{tab:massDs}, \ref{tab:masseta}, the differences of our results compared with the experimental values for the ${1^3}{P_0}$ states, are (-56.960) MeV ($D_0^*(2300)$), 75.361 MeV ($D_{s0}^*{(2317)^ \pm }$) and (-42.680) MeV (${\chi _{c0}}(1P)$), respectively. For the case of the ${1^3}{D_3}$ states, $\Delta M$ equals to the values (-13.366) MeV ({$D_3^*(2750)$}), 3.603 MeV (${D_{s3}^*}(2860)^{\pm }$) and 19.342 MeV ($\psi (3842)$). Note that in the present work, the contribution of spin-orbit interactions is not included. $\Delta M$ for different S-wave states of $D$ mesons have the values 9.381 MeV ($D$), 24.848 MeV ($D^*(2007)$), 0.737 MeV ($D_0(2550)^0$) and 8.899 MeV ($D_1^*(2600)^0$). In the case of $D_s$, we have these differences: 10.253 MeV ($D_s$), 2.738 MeV ($D_s^{*\pm}$) and 9.386 MeV ($D_{s1}^*(2700)^{\pm}$). In Table \ref{tab:masseta}, $\Delta M$ are varied from (-47.452) MeV (for ${\eta _c (1S)}$), (-1.500) MeV (${\eta _c}(2S)$), (-65.072) MeV ($J/\psi (1S)$) to (-34.831) MeV ($\psi (2S)$). Note that the signs of $\Delta M$ are different, where some of them are negative and the other ones have a positive sign, and they depend on the values of $V_0$ in our potential, which is different for each state. For the states of ${1^1}{S_0},{1^3}{S_1}$, we have the same value of $a$ for each meson, since the wave function is dependent on $n, l$ and taken $n=1, l=0$ for these states. We have plotted the wave functions for the states $D$ and $J/\psi$ in Figs. \ref{fig:wave D} and \ref{fig:wave J}, respectively, where they are normalized to one. Thus, for the case of $D$ meson in Fig. \ref{fig:wave D}, we take $n=1$, $l=0$ and $S=0$ for a state with the spectroscopic notation ${1^1}{S_0}$. In Fig. \ref{fig:wave J}, it refers to a case of a vector charmonium state ${1^3}{S_1}$, the $J/\psi$, where one can see that the wave function drops faster than the one of the $D$ meson.

\begin{center}
\begin{table}
\caption{$D$ meson spectra.}
\label{tab:massD}
\begin{tabular}{|p{2.5cm}|p{1.5cm}|p{2.5cm}|p{2.5cm}|p{2.5cm}|p{2.5cm}|}
\hline    
\textbf{${n^{2S + 1}}{L_J},{J^P}$} &  \textbf{\textit{a}}  & \textbf{Our mass (GeV)} & \textbf{Meson} &   \textbf{Exp. mass \cite{PDG:2020}} & \textbf{$\Delta m =$ Our - Exp. (MeV) } \\ \hline  
${1^1}{S_0}, {J^P} = {0^ - }$ & {0.288} & {1.878} & $D$ & 1.869 & {9} \\ \hline
${1^3}{S_1}, {J^P} = {1^ - }$ & {0.288} & {2.031} & ${D^*}(2007)$ & 2.006 & {25} \\ \hline
${2^1}{S_0}, {J^P} = {0^ - }$ & {0.199} & {2.550} & ${D_0}{(2550)^0}$ & 2.549 & {1} \\ \hline
${2^3}{S_1}, {J^P} = {1^ - }$ & {0.199} & {2.636} & $D_1^*{(2600)^0}$ & 2.627 & 9 \\ \hline
${1^3}{P_0}, {J^P} = {0^ + }$ & {0.239} & {2.286} & $D_0^*(2300)$ & 2.343 & {-57} \\ \hline
${1^3}{D_3}, {J^P} = {3^ - }$ & {0.228} & {2.750} & $D_3^*(2750)$ & 2.763 & {-13} \\ \hline
\end{tabular} 
\end{table}
\end{center}

\begin{center}
\begin{table}
\caption{$D_{s}$ meson spectra.}
\label{tab:massDs}
\begin{tabular}{|p{2.5cm}|p{1.5cm}|p{2.5cm}|p{2.5cm}|p{2.5cm}|p{2.5cm}|}
\hline    
\textbf{${n^{2S + 1}}{L_J},{J^P}$} &  \textbf{\textit{a}}  & \textbf{Our mass (GeV)} & \textbf{Meson} &   \textbf{Exp. mass \cite{PDG:2020}} & \textbf{$\Delta m =$ Our - Exp. (MeV)} \\ \hline 
${1^1}{S_0}, {J^P} = {0^ - }$ & {0.343} & {1.978} & $D_s$ & 1.968 & {10} \\ \hline
${1^3}{S_1}, {J^P} = {1^ - }$ & {0.343} & {2.115} & $D_s^{* \pm }$ & 2.112 & {3}  \\ \hline
${2^3}{S_1}, {J^P} = {1^ - }$ & {0.235} & {2.723} & $D_{s1}^*{(2700)^ \pm }$ & 2.714 & {9}  \\ \hline
${1^3}{P_0}, {J^P} = {0^ + }$ & {0.279} & {2.392} & $D_{s0}^*{(2317)^ \pm }$ & 2.317 & {75}  \\ \hline
${1^3}{D_3}, {J^P} = {3^ - }$ & {0.266} & {2.864} & $D_{s3}^*{(2860)^ \pm }$ & 2.860 & {4} \\ \hline
\end{tabular} 
\end{table}
\end{center}

\begin{center}
\begin{table}
\caption{Charmonium meson spectra.}
\label{tab:masseta}
\begin{tabular}{|p{2.5cm}|p{1.5cm}|p{2.5cm}|p{2.5cm}|p{2.5cm}|p{2.5cm}|}
\hline    
\textbf{${n^{2S + 1}}{L_J},{J^P}$} &  \textbf{\textit{a}}  & \textbf{Our mass (GeV)} & \textbf{Meson} &   \textbf{Exp. mass \cite{PDG:2020}} & \textbf{$\Delta m =$ Our - Exp. (MeV)}  
\\ \hline  
${1^1}{S_0}, {J^P} = {0^ - }$ & {0.540} & {2.936} & ${\eta _c}(1S)$ & 2.983 & {-47}  \\ \hline
${1^3}{S_1}, {J^P} = {1^ - }$ & {0.540} & {3.031} & $J/\psi (1S)$ & 3.096 & {-66} \\ \hline
${2^1}{S_0}, {J^P} = {0^ - }$ & {0.362} & {3.635} & ${\eta _c}(2S)$ & 3.637 & {-2}  \\ \hline
${2^3}{S_1}, {J^P} = {1^ - }$ & {0.362} & {3.651} & $\psi (2S)$ & 3.686 & {-35} \\ \hline
${1^3}{P_0}, {J^P} = {0^ + }$ & {0.416} & {3.371} & ${\chi _{c0}}(1P)$ & 3.414 & {-43} \\ \hline
${1^3}{D_3}, {J^P} = {3^ - }$ & {0.390} & {3.861} & ${\psi _3}(3842)$ & 3.842 & {19} \\ \hline
\end{tabular} 
\end{table}
\end{center}

\vspace{-1cm}
\section{Semileptonic Decay Widths }
\label{section:3}
The well-known slope and curvature of the IWF can be calculated through the wave functions of the meson systems, given by \cite{Hassanabadi:2014isa}
\begin{equation}
    \begin{gathered}
  {\rho ^2} = 4\pi {\mu ^2}\int\limits_0^\infty  {{r^4}} {\psi_{n,l}^2 (a,r)}dr, \hfill \\
  C = \frac{2}{3}\pi {\mu ^4}\int\limits_0^\infty  {{r^6}} {\psi_{n,l} ^2 (a,r)}dr, \hfill 
\end{gathered} 
\end{equation}
where $\mu $ is the reduced meson mass. Taking the wave functions obtained above, we tabulate our results in Table \ref{tab:parameters IWF} for the wave functions of the ground states, where we found that the parameters of IWF were sensitive to the variation parameter, $a$. It is obvious that $a$ is dependent on the average of the Hamiltonian and the interactions of the quarks as well. One can expect that by raising the reduced meson masses, the values of the parameters of the IWF are increased \cite{Hassanabadi:2014isa}. The values of these parameters are varied in different models. Blok and Shifman determined the slope parameter as $\rho _{}^2 = 0.7 \pm 0.25$ for the heavy meson system and also obtained a range of 0.35 to 1.15 for different Borel parameters \cite{Blok:1992fc} by taking the Bjorken sum rule method. Moreover, using the Lattice QCD, $\rho _{}^2 = 1.2_{ - 3}^{ + 7}$ was found by Booth et al. \cite{UKQCD:1993lfy} for the $D$ meson. In the previous works, we have calculated the slope and curvature, obtained $\rho _{}^2 = 0.62$, $C = 0.09$  \cite{Hassanabadi:2014isa} and  $\rho _{}^2 = 0.69$, $C = {\text{0}}{\text{.14}}$ \cite{Rahmani:2017vbg} for the $D$ meson, where the differences compared to our present ones are {0.072} \cite{Hassanabadi:2014isa} and {0.002} \cite{Rahmani:2017vbg} for the slope quantity. For the curvature, the differences of our values with these previous ones are about 0.043 \cite{Hassanabadi:2014isa} and {0.007} \cite{Rahmani:2017vbg}. In Ref. \cite{Rahmani:2017vbg}, we employed a generalized quantum isotonic oscillator potential along with a Gaussian-type wave function. In Ref. \cite{Hassanabadi:2014isa}, we have obtained Airy functions for the wave function of the heavy mesons. But in the present work, our formalism are different that the harmonic oscillator wave functions and the Yukawa term for the interactions are taken into account. Since we are dealing with the semileptonic decays of the $D^0$ and $D_s$ mesons in the next steps, we should obtain firstly the parameters of the IWF for the ground states of charmed mesons, which are applied to calculate the form factors. It is worthwhile to mention that the IWF is related to the ground state wave functions \cite{Blok:1992fc}.

\begin{center}
\begin{table}[ht]
\caption{ ${\rho ^2}$ and $C$ for the charmed mesons.}
\label{tab:parameters IWF}
\begin{tabularx}{\textwidth}{|X|X|X|X|X|} 
\hline    

 \textbf{Meson}  & \textbf{${\rho ^2}$ (this work)} &  \textbf{${\rho ^2}$ (others)} & \textbf{$C$ (this work)} & \textbf{$C$ (others)} \\ \hline  
   
  $D$  & {0.69} & 0.68 \cite{Roy:2012ng} & {0.13} & 0.11 \cite{Roy:2012ng} \\ \hline
  $D_s$  & {0.86} & 0.79 \cite{Roy:2012ng} & {0.21} & 0.28 \cite{Roy:2012ng} \\ \hline
  ${\eta _c}(1S)$  &{1.54} & - & {0.66} & - \\ \hline

\end{tabularx} 
\end{table}
\end{center}

\begin{figure}
\centering
\includegraphics[width=0.48\linewidth]{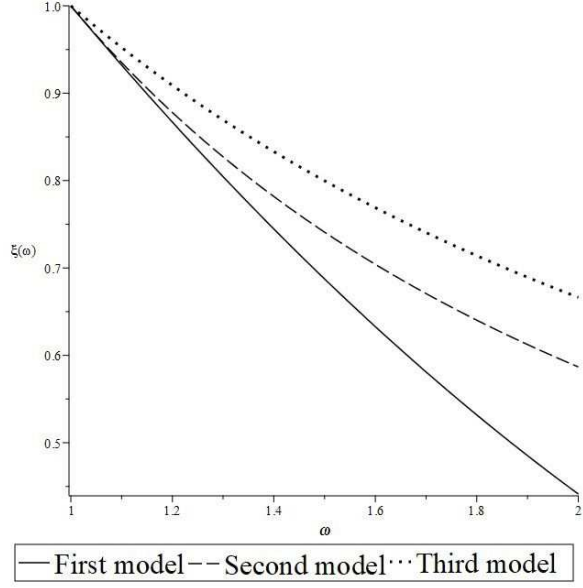}
\caption{ Different IWFs versus $\omega$ for the transition ${D^0} \to {K^ - }{\ell ^ + }{\nu _\ell }$.}
\label{fig:IWFs}
\end{figure}

\vspace{-1cm}
 Next, the semileptonic decay width of ${D^0} \to {K^ - }{l^ + }\upsilon $  can be obtained through the following relations \cite{Zhang:2018jtm, Hwang:1998ph},
\begin{equation}
    \begin{gathered}
  \frac{{d\Gamma ({D^0} \to {K^ - }{l^ + }\nu )}}{{d{q^2}}} = \frac{{G_F^2}}{{24{\pi ^3}}}|{V_{cs}}{|^2}p({q^2}){(1 - \frac{{m_l^2}}{{{q^2}}})^2} \times  \hfill \\
  [{(p({q^2}))^2}(1 + \frac{1}{2}\frac{{m_l^2}}{{{q^2}}})|{F_1}({q^2}){|^2} + M_D^2{(1 - \frac{{M_K^2}}{{M_D^2}})^2}\frac{3}{8}\frac{{m_l^2}}{{{q^2}}}|{F_0}({q^2}){|^2}], \hfill 
  \label{eq:DtoK}
\end{gathered} 
\end{equation}
in terms of the hadronic form factors  ${F_1}({q^2})$ and ${F_0}({q^2})$, where $M_D$ and $M_K$ are the masses of $D$ and $K$ mesons and $m_l$ is the lepton mass. We have taken the $D$ meson mass from Table \ref{tab:massD}. Besides, $G_F$  is the Fermi coupling constant and $p({q^2})$ momentum of the final meson in the initial meson rest frame, which can be written as,
\begin{equation}
    |\vec p({q^2})| = \frac{{\sqrt {\lambda (M_D^2,M_K^2,{q^2})} }}{{2{M_D}}},
\end{equation}
in terms of
\begin{equation}
    \lambda (x,y,z) = {x^2} + {y^2} + {z^2} - 2xy - 2yz - 2zx.
\end{equation}
Using the heavy quark effective theory, the relations between the form factors and the IWF are given by \cite{Neubert:1991xw}
\begin{equation}
    \begin{gathered}
  {F_1}({q^2}) = \frac{{{M_D} + {M_K}}}{{2\sqrt {{M_D}{M_K}} }}\xi (\omega ), \hfill \\
  {F_0}({q^2}) = \frac{{2\sqrt {{M_D}{M_K}} }}{{{M_D} + {M_K}}}\frac{{\omega  + 1}}{2}\xi (\omega ), \hfill \\ 
\end{gathered} 
\end{equation}
with
\begin{equation}
    \omega  = \frac{{M_D^2 + M_K^2 - {q^2}}}{{2{M_D}{M_K}}},
\end{equation}
where $\xi (\omega )$ is the IWF. Integrating Eq. (\ref{eq:DtoK}) over the range of $m_l^2 \leqslant {q^2} \leqslant {({M_D} - {M_K})^2}$, we can calculate the branching ratio for this transition. We use three different models for the IWF. First, using the well-known IWF, we firstly take the following series \cite{Hassanabadi:2014isa, Rahmani:2017vbg},
 \begin{equation}
     \xi (\omega ) = 1 - {\rho ^2}(\omega  - 1) + C{(\omega  - 1)^2} + ..., 
     \label{IWF:expand}
 \end{equation}
 which is placed at a small recoil point. The experimentalists much concern the straightforward procedure of expanding form factors in the vicinity of zero recoil point, where $\omega$ has the value of one. $\xi (\omega)$ in Eq. (\ref{IWF:expand}) is based on two parameters $\rho^2$ and $C$. Taking the parameters of the IWF, $\rho^2$ and $C$ from Table \ref{tab:parameters IWF}, we can obtain the transition form factors. The second model is adopted the oscillator parameterization \cite{Neubert:1991xw, UKQCD:1993lfy},
 \begin{equation}
     \xi (\omega ) = \frac{2}{{\omega  + 1}}\exp \left( { - (2{\rho ^2} - 1)\frac{{\omega  - 1}}{{\omega  + 1}}} \right),
     \label{IWF:exp}
 \end{equation}
 which is based on a simple relativistic oscillator model \cite{Neubert:1991xw} and depends on the dimensionless parameter of the slope $\rho^2$.
 The third model is related to the zeroth spherical Bessel function, mesonic wave function and energy $E$ in the ${D^0} \to {K^ - }{l^ + }\nu $ transitions \cite{Coleman:2000gu},
 \begin{equation}
     \xi (\omega ) = \frac{2}{{\omega  + 1}}\left\langle {{j_0}(2Er\sqrt {\frac{{\omega  - 1}}{{\omega  + 1}}} )} \right\rangle, 
     \label{IWF:wave function}
 \end{equation}
 where the notation $\left\langle {{j_0}(2Er\sqrt {\frac{{\omega  - 1}}{{\omega  + 1}}} )} \right\rangle $ is the expectation value of ${{j_0}(2Er\sqrt {\frac{{\omega  - 1}}{{\omega  + 1}}} )}$, obtained by the wave function of the system as given in Eq. (\ref{wave function}) with $n=1$ and $l=0$. Using Eq. (\ref{energy}), we have also obtained $E_{1,0}$ = -0.014 GeV for the $D$ meson and $E_{1,0}$ = -0.074 GeV for the $D_s$ meson. We have taken the variation parameter of the wave function for the $D$ and $D_s$ mesons from what we obtained in Tables \ref{tab:massD} and \ref{tab:massDs}. We have plotted the behaviours of the IWFs in Fig. \ref{fig:IWFs}. As one expects from the IWFs, at $\omega = 1$, it gives $\xi (\omega ) =$ 1 due to the current conservation, all of which start from one and then reduce by increasing $\omega$. 
 
 The CKM matrix element should be changed to  $V_{cd}$ in the case ${D^0} \to {\pi ^ - }{l^ + }\nu $. We have taken the input values of the masses, CKM matrix elements, and lifetimes as 
 $ \left| {{V_{cs}}} \right| = 0.987$, $|{V_{cd}}| = 0.221,$ ${M_\eta } = 0.547862$ GeV, ${M_{{K^ \pm }}} = 0.493677$ GeV, ${M_{{K^0}}} = 0.497611$ GeV, ${M_{{\pi ^ \pm }}} = 0.13957039$ GeV, ${m_e} = 0.510 \times {10^{ - 3}}$ GeV, ${m_\mu } = 0.1056$ GeV, ${\tau _{{D^0}}} = 4.101 \times {10^{ - 13}}s$ and ${\tau _{{D_s}}} = 5.04 \times {10^{ - 13}}s$ \cite{PDG:2020}. The branching fractions of the semileptonic decay widths of $D$ have been obtained as \cite{PDG:2020},
  \par
  $\left\{ \begin{gathered}
  BR({D^0} \to {K^ - }{e^ + }{\nu _e}) = (3.541 \pm 0.034)\%  \hfill \\
  BR({D^0} \to {K^ - }{\mu ^ + }{\nu _\mu }) = (3.41 \pm 0.04)\%  \hfill \\ 
\end{gathered}  \right.$, $\left\{ \begin{gathered}
  BR({D^0} \to {\pi ^ - }{e^ + }{\nu _e}) = (2.91 \pm 0.04) \times {10^{ - 3}} \hfill \\
  BR({D^0} \to {\pi ^ - }{\mu ^ + }{\nu _\mu }) = (2.67 \pm 0.12) \times {10^{ - 3}} \hfill \\ 
\end{gathered}  \right.$,
 for different decay channels of the leptons \cite{PDG:2020}. Our results in Table \ref{tab:BRDtoK} are compatible with them, where the deviations of our results to theirs for the case of ${D^0} \to {K^ - }{e^ + }\nu_e $  are 0.331, 0.084 and 0.117, regarding to the first, second and third models respectively. In the case of ${D^0} \to {K^ - }{\mu ^ + }{\nu _\mu }$, the deviations are 0.318, 0.076 and 0.125, respectively. We can see that the second and third models lead to better results for the semileptonic decay widths of the transitions ${D^0} \to {K^ - }$. The results of third model for $BR({D^0} \to {\pi ^ - }{e^ + }{\nu _e})$ and $BR$(${D^0} \to {\pi^ - }{ \mu ^ +}{\nu _\mu }$) are closer to the experimental values with the deviations of about 0.444 and 0.407, respectively. We also find that the ratio $BR({D^0} \to {\pi ^ - }{e^ + }{\nu _e})/BR({D^0} \to {K^ - }{e^ + }{\nu _e})$  for the third model is given as 0.041, which is near one half of the result $0.085 \pm 0.007$ in Ref. \cite{PDG:2020}. The ratio of  $BR({D^0} \to {\pi ^ - }{\mu ^ + }{\nu _\mu })/BR({D^0} \to {K^ - }{\mu ^ + }{\nu _\mu })$ was given as $0.074 \pm 0.008$ in Ref. \cite{PDG:2020}, where ours is 0.041 from the third model. Based on Eqs. (\ref{IWF:expand}, \ref{IWF:exp}) for the first and second models, respectively, we have $BR({D^0} \to {\pi ^ - }{\mu ^ + }{\nu _\mu })/BR({D^0} \to {K^ - }{\mu ^ + }{\nu _\mu })$ = 0.393 and 0.031, respectively.
 
\begin{center}
\begin{table}[ht]
\caption{Branching fractions for ${D^0} \to {K^ - }{l^ + }\nu $  and ${D^0} \to {\pi ^ - }{l^ + }\nu $ (in \%).}
\label{tab:BRDtoK}
\begin{tabularx}{\textwidth}{|X|X|X|X|X|X|} 
\hline    
\textbf{Decay}  & \textbf{This work (Eq. \ref{IWF:expand})} & \textbf{This work (Eq. \ref{IWF:exp})} & \textbf{This work (Eq. \ref{IWF:wave function})} & \textbf{Others} & \textbf{Exp. \cite{PDG:2020}} \\  \hline  
${D^0} \to {K^ - }{e^ + }{\nu _e}$ & {2.369} & {3.243} & {3.956} & 3.49, 4.78 \cite{Hwang:1998ph}; 3.2 \cite{Wu:2006rd} &  $3.541 \pm 0.034$ 
 \\  \hline
${D^0} \to {K^ - }{\mu ^ + }{\nu _\mu }$ & {2.326} & {3.152} & {3.835} & 3.38, 4.67 \cite{Hwang:1998ph}; 3.15 \cite{Wu:2006rd} & $ 3.41 \pm 0.04$
 \\  \hline
${D^0} \to {\pi ^ - }{e^ + }{\nu _e}$ & {1.018} & {0.099} & {0.162} & 0.292, 0.594 \cite{Hwang:1998ph}; 0.278 \cite{Wu:2006rd} &  0.291  
\\  \hline
${D^0} \to {\pi ^ - }{\mu ^ + }{\nu _\mu }$ & {0.914} & {0.097} & {0.158} & 0.288, 0.586 \cite{Hwang:1998ph}; 0.275 \cite{Wu:2006rd}  & 0.267 
\\  \hline
\end{tabularx} 
\end{table}
\end{center}
\par

\vspace{-1cm}
We also obtain the semileptonic decay widths of $D_s$ using the same approach, Eqs. (\ref{eq:DtoK}) to (\ref{IWF:wave function}), with the replacement of $|{V_{cs}}|$  by $|{V_{cd}}|$  for $D_s^{} \to {K^0}{e^ + }\nu_e $. For the transition of ${D_s} \to \eta {l^ + }\nu_l $ , we still use $|{V_{cs}}| $ in Eq. (\ref{eq:DtoK}). We take the meson mass of $D_s$ from our obtained value in Table \ref{tab:massDs}. In Table \ref{tab:D_stoK}, we show our results for the semileptonic decay widths of $D_s$ compared with the experimental ones. The deviations of our values are 0.610, 0.357 and 0.058 for the first, second and third models, respectively, regarding the semileptonic decay width $D_s^{} \to {K^0}$. The results of first model are closer to the experimental ones with the deviations about {0.210 and 0.269 }for the decays $D_s^{} \to \eta {\mu ^ + }{\nu _\mu }$  and  $D_s^{} \to \eta {e^ + }{\nu _e}$, respectively. Azizi et al. calculated the branching fraction of $D_s$, obtained as $BR({D_s} \to \eta l\nu ) = 3.15 \pm 0.97\,\,\% $ via light cone QCD sum rules \cite{Azizi:2010zj}, which is consistent with ours within the uncertainties.

\begin{center}
\begin{table}[ht]
\caption{Branching fractions for ${D_s} \to {K^0}{l^ + }\nu $  and  ${D_s} \to \eta {l^ + }\nu $.}
\label{tab:D_stoK}
\begin{tabularx}{\textwidth}{|X|X|X|X|X|} 
\hline    
\textbf{Decay}  & \textbf{This work (Eq. \ref{IWF:expand})} & \textbf{This work (Eq. \ref{IWF:exp})} & \textbf{This work (Eq. \ref{IWF:wave function})} & \textbf{Exp. \cite{PDG:2020}} \\  \hline      
   $D_s^{} \to {K^0}{e^ + }{\nu _e}$ & {$({\text{1}}{\text{.327}}) \times {10^{ - 3}}$} & { $({\text{2}}{\text{.186}}) \times {10^{ - 3}}$}  & {$({\text{3}}{\text{.203}}) \times {10^{ - 3}}$} & $ (3.4 \pm 0.4) \times {10^{ - 3}}$\\  \hline 
 $D_s^{} \to \eta {e^ + }{\nu _e}$ & {2.944} \%  & {4.278} \% & {6.008} \% & $({\text{2}}{\text{.32}} \pm {\text{0}}{\text{.08}})\% $\\  \hline
  $D_s^{} \to \eta {\mu ^ + }{\nu _\mu }$ & {2.905} \%  & {4.174} \% & {5.837} \% & $({\text{2}}{\text{.4}} \pm {\text{0}}{\text{.5}})\%  $\\  \hline
\end{tabularx} 
\end{table}
\end{center}

\par
\begin{figure}
\centering
\includegraphics[width=0.48\linewidth]{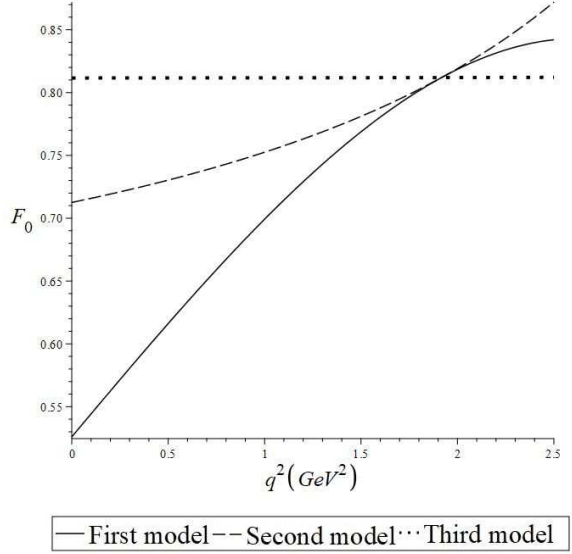}
\caption{$F_0(q^2)$ versus $q^2$ for ${D^0} \to {K^ - }{l^ + }\nu_l $.}
\label{fig:F_0}
\end{figure}
\begin{figure}
\centering
\includegraphics[width=0.48\linewidth]{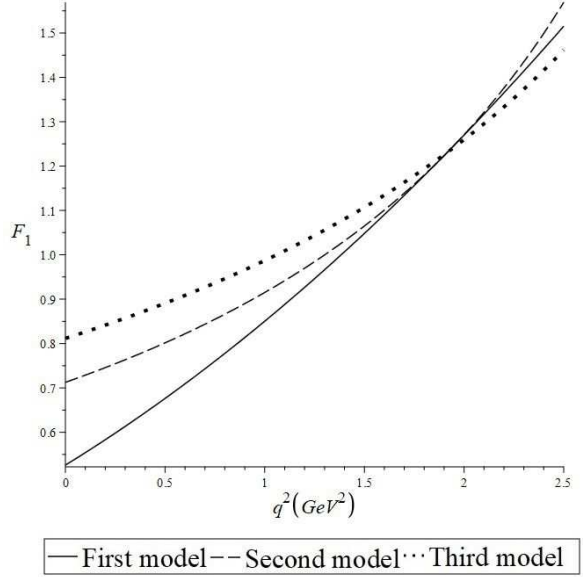}
\caption{$F_1(q^2)$ versus $q^2$ for ${D^0} \to {K^ - }{l^ + }\nu_l $.}
\label{fig:F_1}
\end{figure}

We plot the form factors $F_0(q^2)$ and ${F_1}({q^2})$ for the decay ${D^0} \to {K^ - }{l^ + }\nu_l$ in Figs. \ref{fig:F_0} and \ref{fig:F_1}, respectively. As one can be seen from these figures, when four momentum transfer equals to zero ($q^2=0$), two form factors should be equal and their values are between 0.5 and 1 for three considered models regarding Eqs. (\ref{IWF:expand}) to (\ref{IWF:wave function}). We obtain ${F_0}({q^2} = 0) = {F_1}({q^2} = 0) = 0.{\text{526}}$, ${F_0}({q^2} = 0) = {F_1}({q^2} = 0) = 0.{\text{713}}$, ${F_0}({q^2} = 0) = {F_1}({q^2} = 0) = 0.{\text{812}}$ using Eqs. (\ref{IWF:expand}), (\ref{IWF:exp}) and (\ref{IWF:wave function}), respectively, for transitions of ${D^0} \to {K^ - }{l^ + }\nu_l$. In this case, 0.716 had been reported for the form factors of ${D^0} \to {K^ - }$ at $q^2=0$ by Faustov et al. \cite{Faustov:2019mqr}. The growing of $F_1(q^2)$ are sharper than the case of $F_0(q^2)$ by increasing the momenta, which agree well with the results of  ${D^0} \to {K^ - }$ transition in Ref. \cite{Lubicz:2017syv}, where the authors obtained $F_0(q^2)$ and $F_1(q^2)$ versus different values of $q^2$ in the range of 0 to 1.8846 $(GeV^2)$. The behaviors of the second and third models are more in agreement with the predictions of lattice QCD results for the form factors of ${D^0} \to {K^ - }$ semileptonic transitions \cite{Lubicz:2017syv}.

\par
\begin{figure}[ht]
  \subfloat[${D^0} \to {\pi ^ - }{\mu ^ + }{\nu _\mu }$.]{
	\begin{minipage}[c][1\width]{
	   0.3\textwidth}
	   \centering
	   \includegraphics[width=1\textwidth]{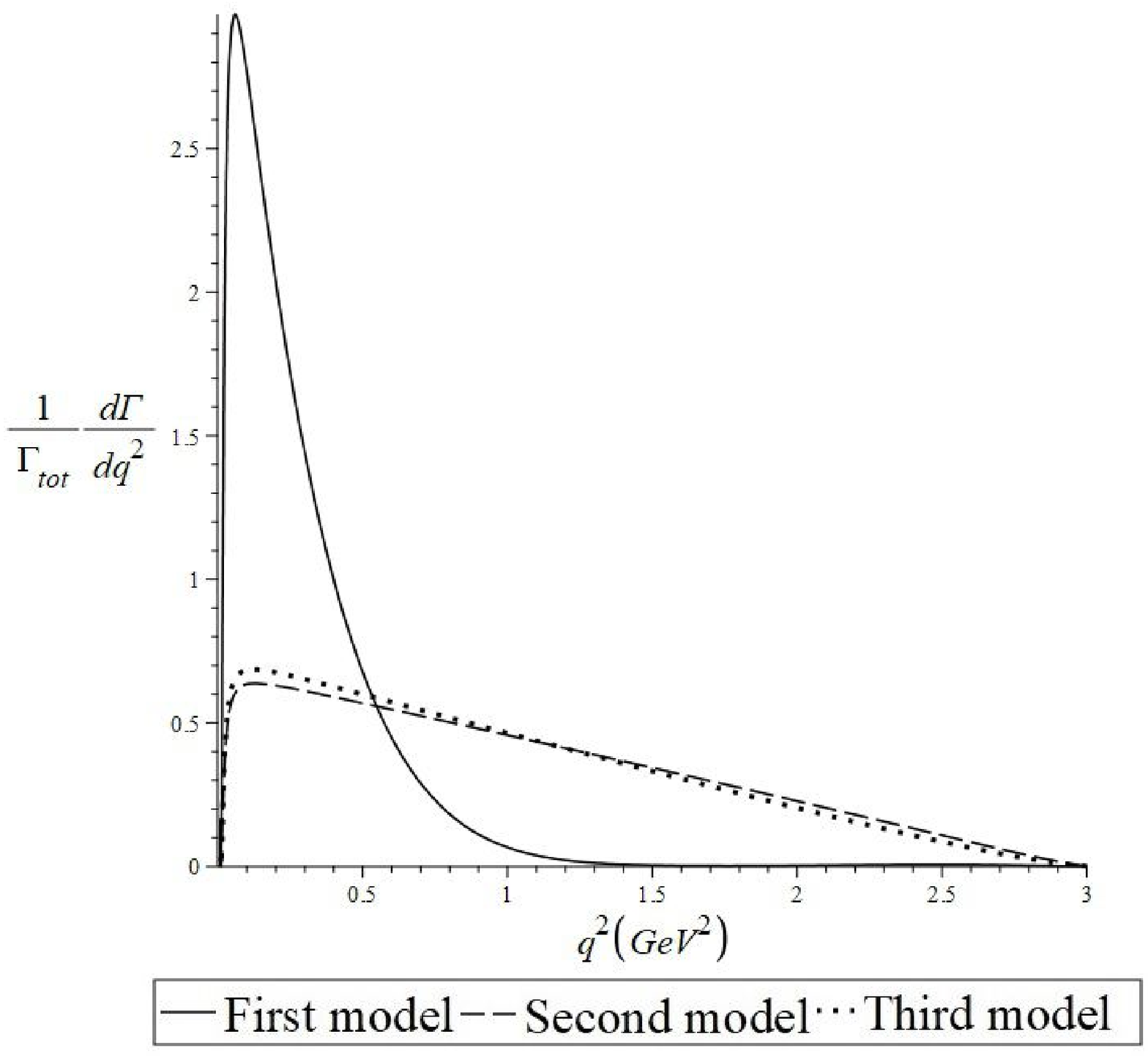}
    \label{D to pion meon}
	\end{minipage}}
  \subfloat[${D^0} \to {\pi ^ - }{e^ + }{\nu _e}$.]{
	\begin{minipage}[c][1\width]{
	   0.3\textwidth}
	   \centering
	   \includegraphics[width=1.1\textwidth]{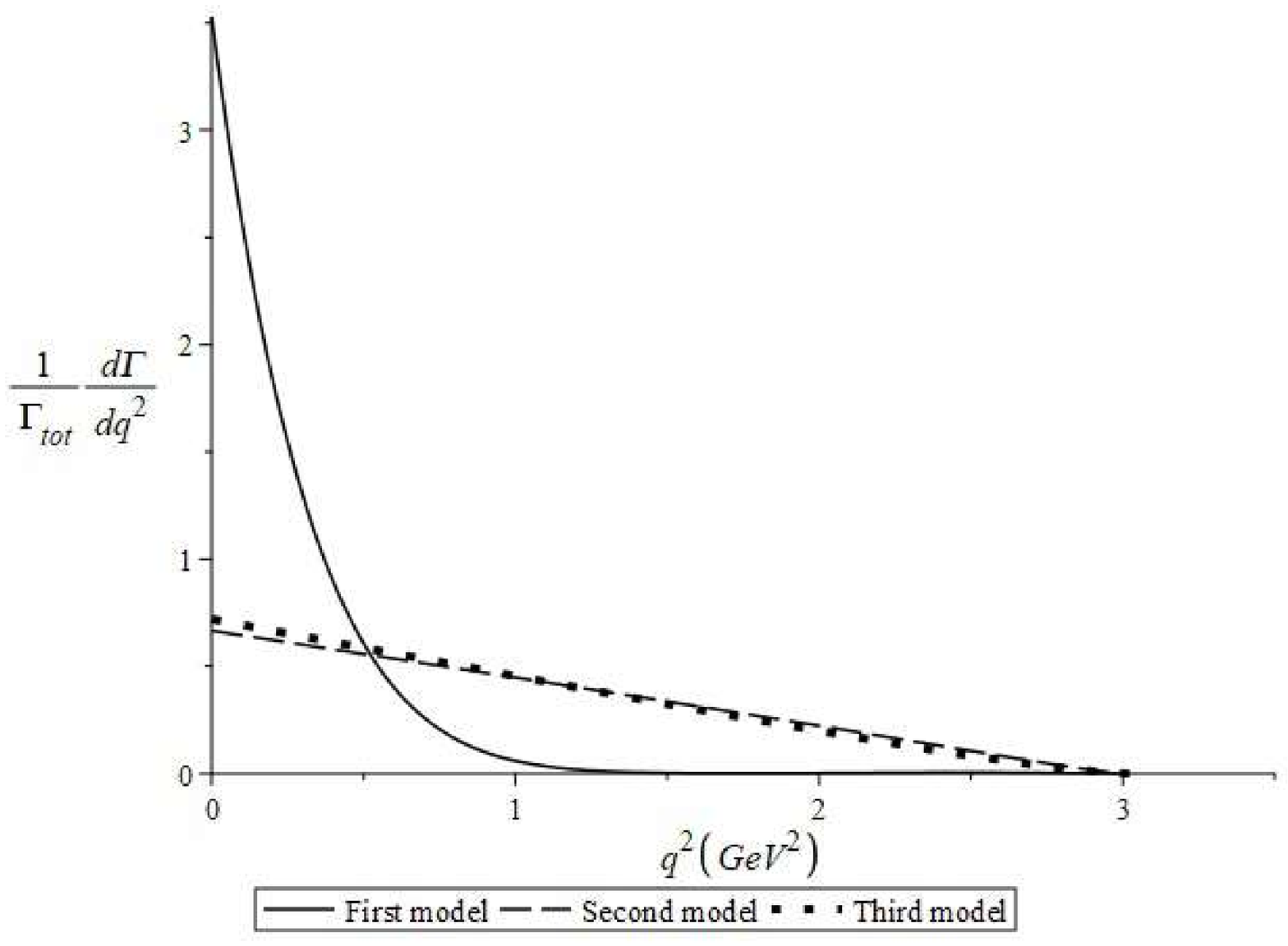}
    \label{fig:D to pion electron}
	\end{minipage}}
\hfill
  \subfloat[${D^0} \to {K ^ - }{e^ + }{\nu _e}$.]{
	\begin{minipage}[c][0.85\width]{
	   0.3\textwidth}
	   \centering
	   \includegraphics[width=1.1\textwidth]{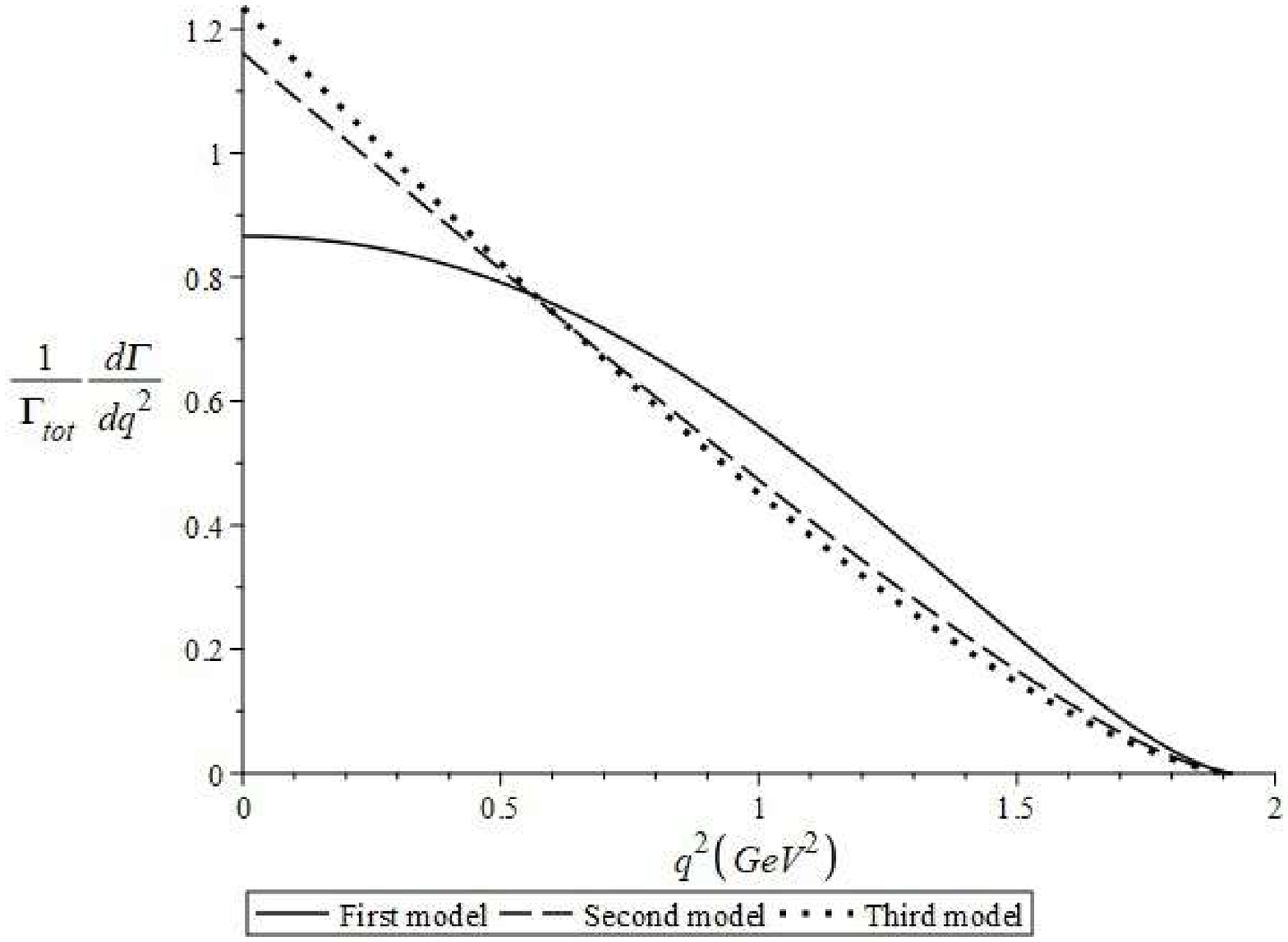}
    \label{fig:D to K electron}
	\end{minipage}}
  \subfloat[ ${D^0} \to {K ^ - }{\mu^ + }{\nu _\mu}$.]{
	\begin{minipage}[c][1\width]{
	   0.3\textwidth}
	   \centering
	   \includegraphics[width=1.1\textwidth]{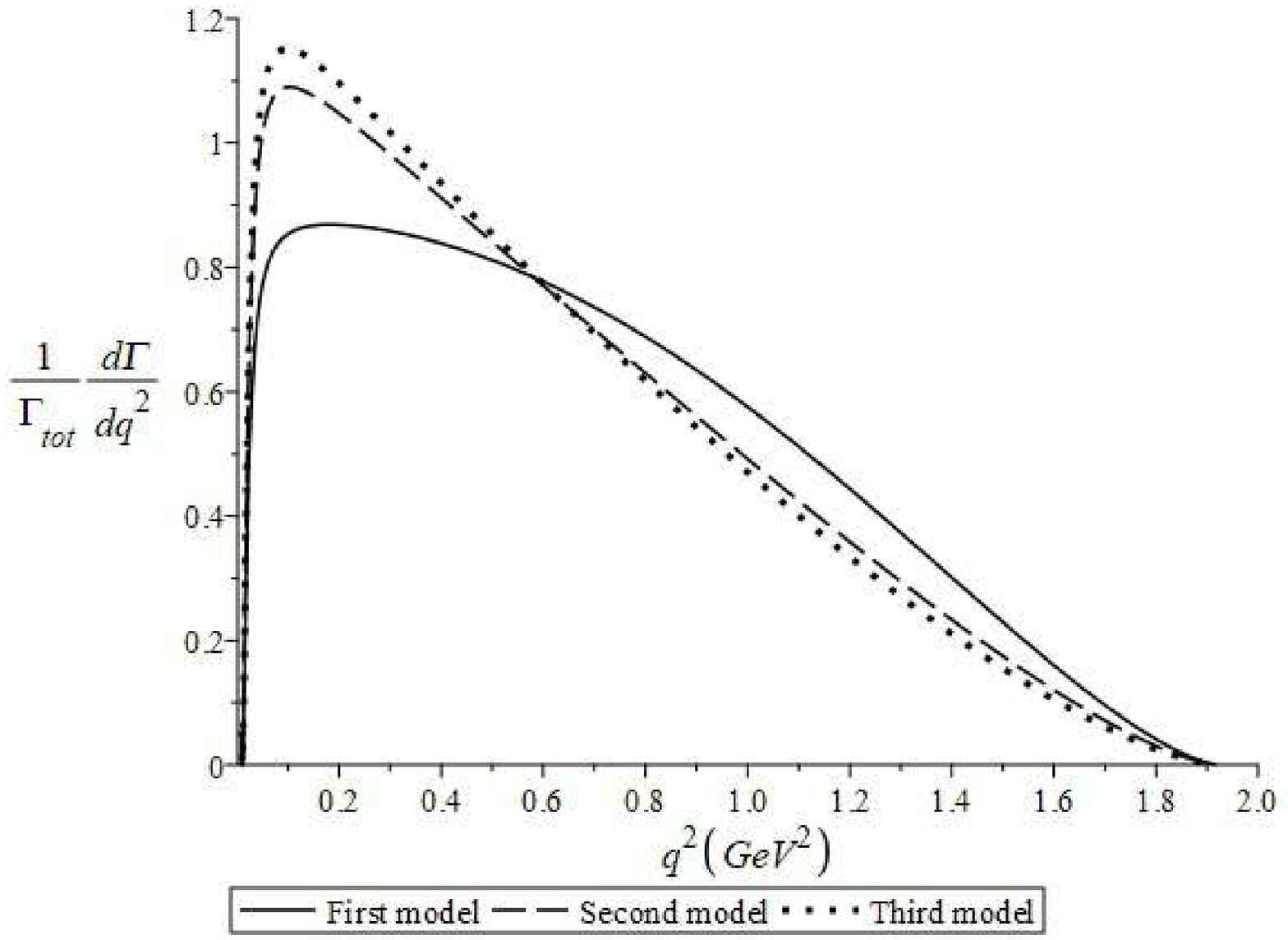}
	\end{minipage}}
\caption{Behaviours of the differential semileptonic decay widths versus $q^2$ for the different channels and three models of the IWFs corresponding to the $D^0$ meson decays.}
\label{subfig:fourfigs}
\end{figure}
\begin{figure}[ht]
  \subfloat[${D_s} \to {K ^ 0 }{e^ + }{\nu _e}$.]{
	\begin{minipage}[c][1\width]{
	   0.3\textwidth}
	   \centering
	   \includegraphics[width=1.1\textwidth]{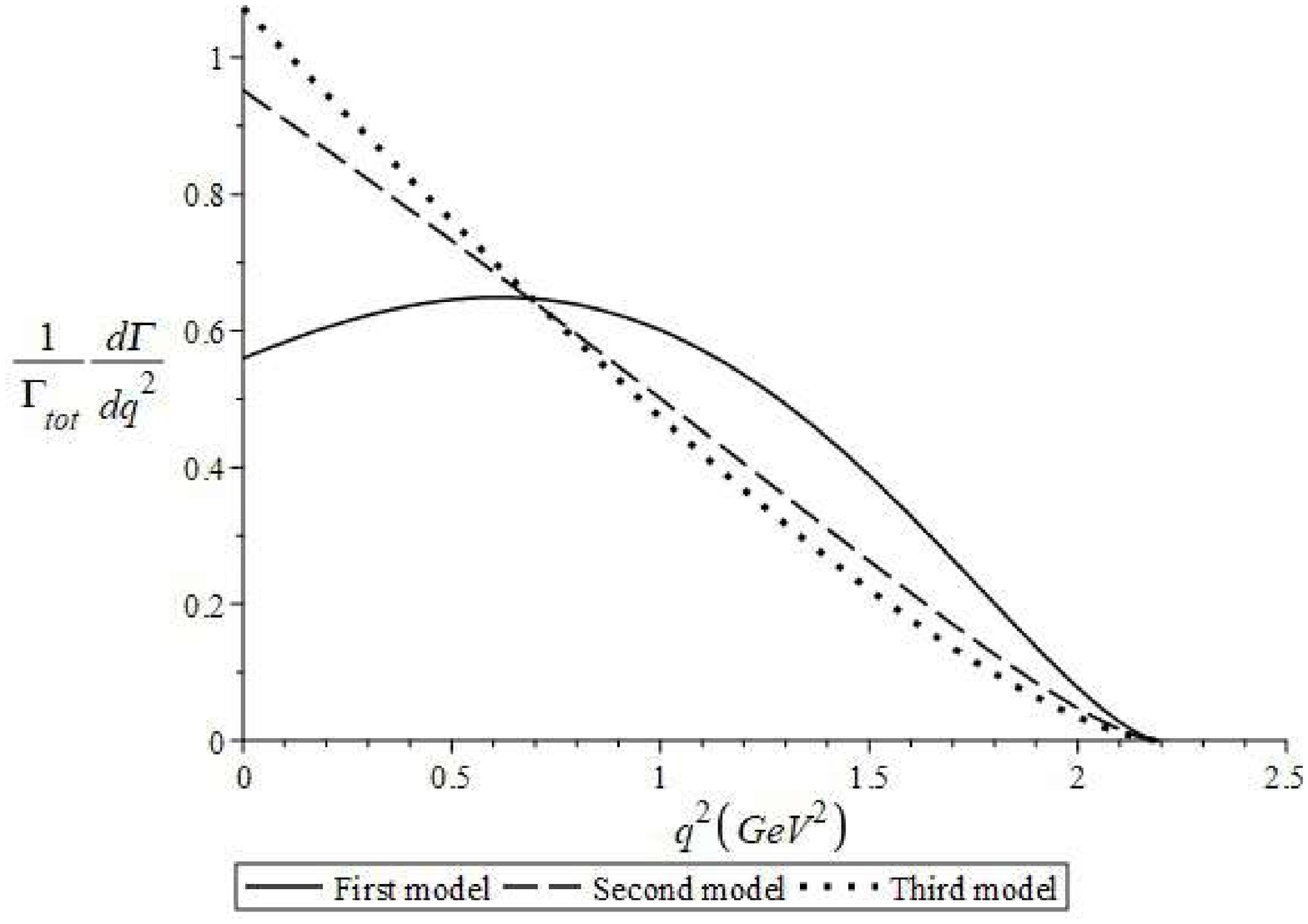}
    \label{fig:Ds to K e}
	\end{minipage}}
 \hfill 	
  \subfloat[${D_s} \to {\eta }{e^ + }{\nu _e}$.]{
	\begin{minipage}[c][1\width]{
	   0.3\textwidth}
	   \centering
	   \includegraphics[width=1.1\textwidth]{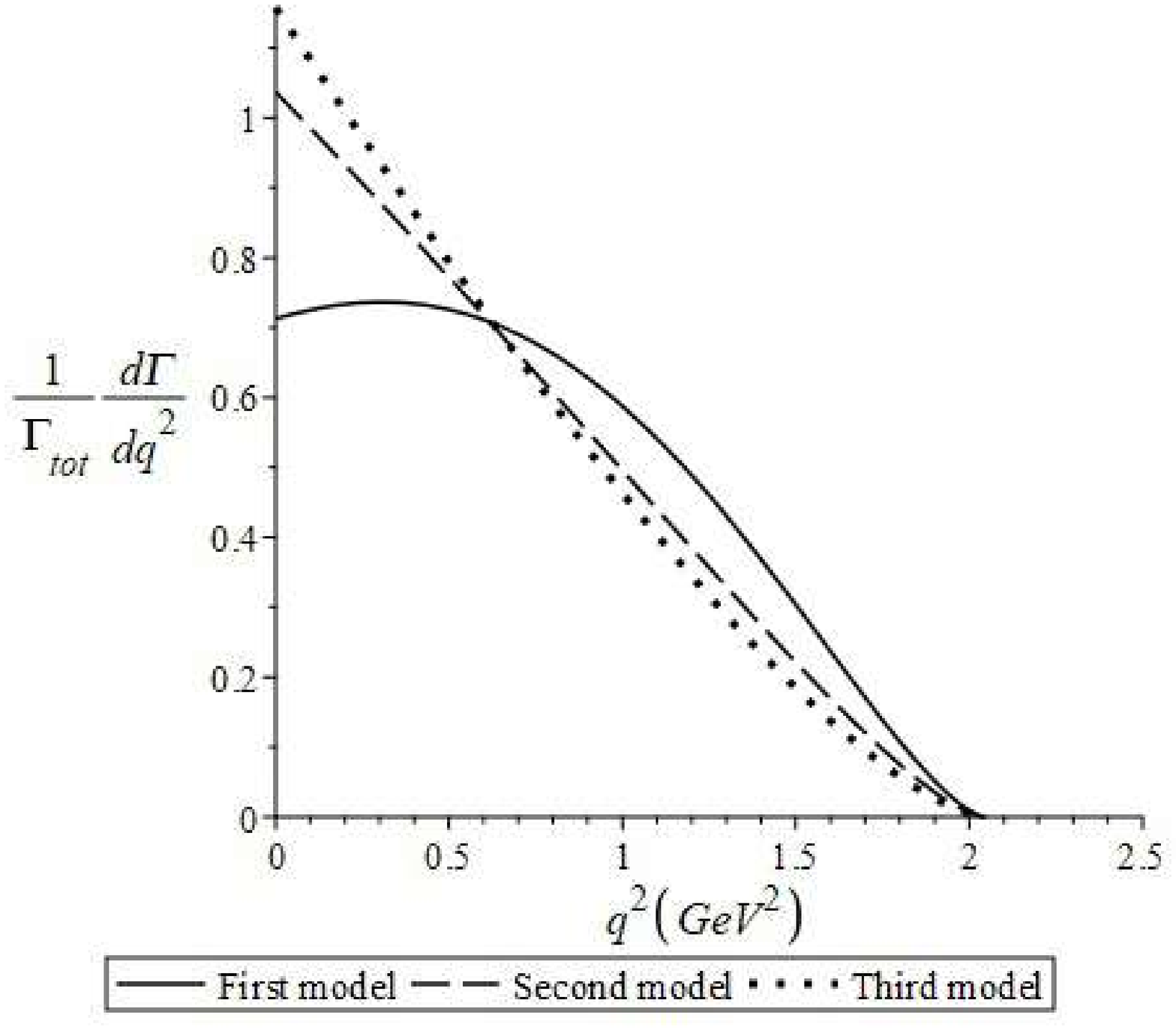}
    \label{fig:Ds to eta e}
	\end{minipage}}
 \hfill	
  \subfloat[${D_s} \to {\eta }{\mu^ + }{\nu _\mu{}{}}$.]{
	\begin{minipage}[c][1\width]{
	   0.3\textwidth}
	   \centering
	   \includegraphics[width=1.1\textwidth]{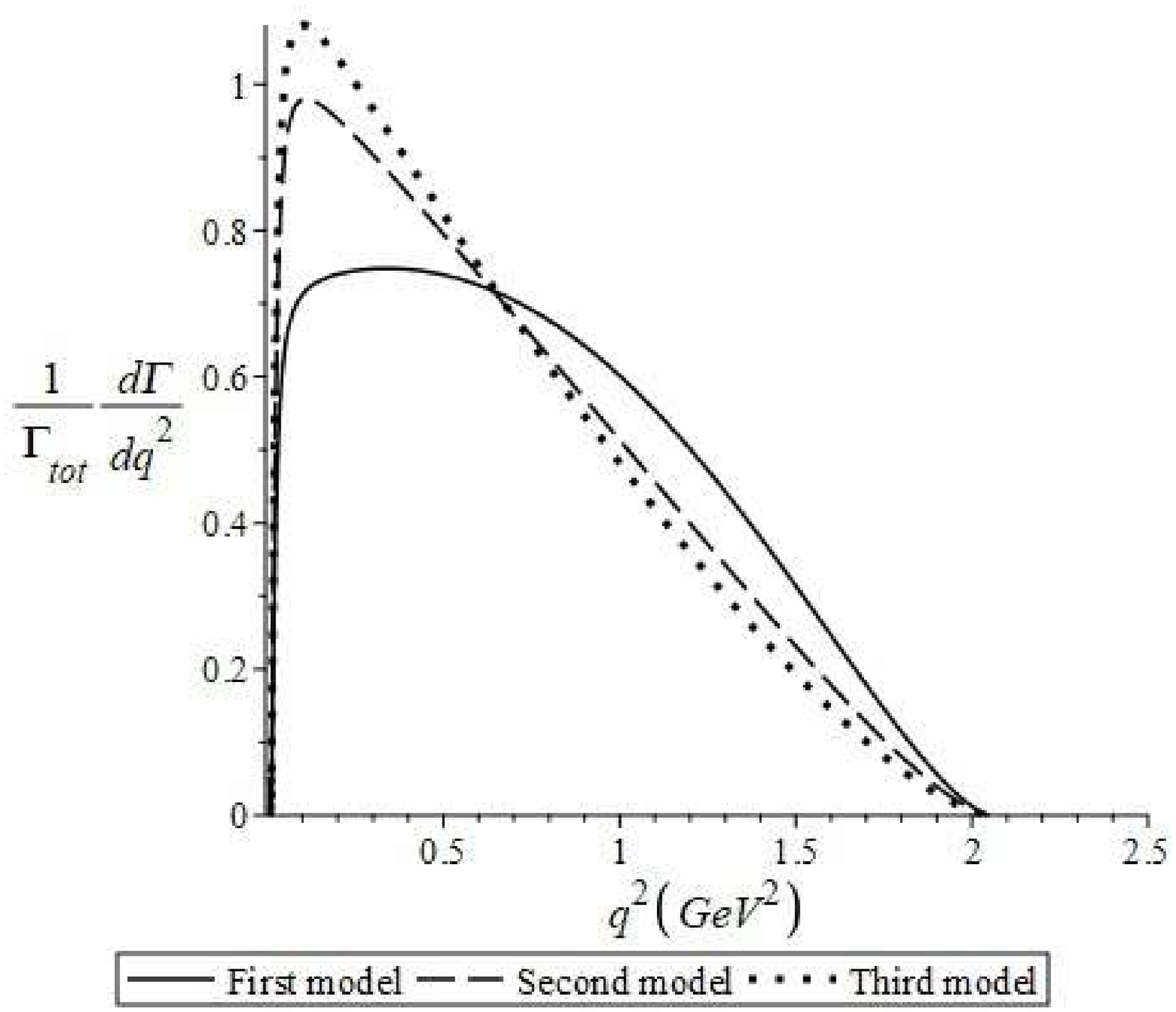}
	\end{minipage}}
\caption{$\frac{{d\Gamma }}{{d{q^2}}}(GeV)$ versus ${q^2}(Ge{V^2})$ for the different channels and three models of the IWFs corresponding to the $D_s$ meson decays.}
\label{subfig2:fourfigs}
\end{figure}

We also plot the behaviors of the semileptonic decay width of $\frac{{d\Gamma }}{{d{q^2}}}(GeV)$  versus ${q^2}(Ge{V^2})$  for the different channels and three models of the IWF in Fig. \ref{subfig:fourfigs}, which corresponds to Table \ref{tab:BRDtoK} for the $D^0$ meson decays, and Fig. \ref{subfig2:fourfigs}, which is for the $D_s$ meson decays regarding to Table \ref{tab:D_stoK}. As one can be seen from Fig. \ref{D to pion meon}, for the transition ${D^0} \to {\pi ^ - }{\mu ^ + }{\nu _\mu }$, the shape of differential semileptonic decay widths for three models start from zero and have peaks at around $q^2$ = 0.059 $GeV^2$, 0.127 $GeV^2$, 0.121 $GeV^2$ for three models accordingly, and then reduce by increasing momenta. In Fig. \ref{D to pion meon}, one can get ${\Gamma _{tot}({D^0} \to {\pi ^ - }{\mu ^ + }{\nu _\mu })} = {\text{1}}{\text{.466}} \times {10^{ - 14}}$ GeV, ${\text{1}}{\text{.562}} \times {10^{ - 15}}$ GeV and ${\text{2}}{\text{.542}} \times {10^{ - 15}}$ GeV regarding the first model, second model and third one, consecutively. We also obtain $\Gamma({D^0} \to {K ^ - }{e^ + }{\nu _e}) = {\text{3}}{\text{.801}} \times {10^{ - 14}}$ GeV for the first model, $5.203 \times 10^{-14}$ GeV for the second model and $6.347 \times 10 ^{-14}$ GeV for the third model.
In the case of ${D^0} \to {\pi ^ - }{e^ + }{\nu _e}$ in Fig. \ref{fig:D to pion electron}, the differential semileptonic decay widths have the nonzero values at the vicinity of zero momentum transfer and then diminish, where the ones of Figs. \ref{fig:D to K electron}, \ref{fig:Ds to K e} and \ref{fig:Ds to eta e} have similar behaviours. In fact, the behaviours of differential semileptonic decay widths of ${D^0} \to {K ^ - }{e^ + }{\nu _e}$ in Fig. \ref{fig:D to K electron}, ${D_s} \to {K ^ 0 }{e^ + }{\nu _e}$ in Fig. \ref{fig:Ds to K e} and $D_s \to {\eta }{e^ + }{\nu _e}$ in Fig. \ref{fig:Ds to eta e} are the same.

Furthermore, we calculate the semileptonic branching ratios of ${D^0} \to {K^ - }{l^ + }\nu $  and ${D^0} \to {\pi ^ - }{l^ + }\nu $   using the parameterization of form factors as the dipole form \cite{Wirbel:1985ji},
\begin{equation}
    {F_1}({q^2}) = \frac{{{F_1}(0)}}{{1 - \frac{{{q^2}}}{{M_{pole}^2}}}},
    {F_0}({q^2}) = \frac{{{F_0}(0)}}{{1 - \frac{{{q^2}}}{{M_{pole}^2}}}},
    \label{eq:dipole}
\end{equation}
where we take the nearest pole dominance for each form factor. For instance, the form factor  ${F_1}({q^2})$ in the $D \to K$ transition is expected to be dominated by the $1^-$ pole \cite{Wirbel:1985ji}. Thus, we take $M_{pole}$ from the $1^-$ states in Table \ref{tab:massD} for the case of $F_1(q^2)$. For $F_0(q^2)$,  the position of $0^+$ pole is needed \cite{Wirbel:1985ji}. Hence, we use the mass value of $0^+$ states for this case. Regarding the PDG values of the form factors at $q^2 = 0$, ${F_0}(0) = {F_1}(0) = 0.637$ and ${F_0}(0) = {F_1}(0) = 0.736$ \cite{PDG:2020} for ${D^0} \to {\pi ^ - }{l^ + }{\nu _l} $ and ${D^0} \to {K^ - }{l^ + }{\nu _l}$, respectively, these results have been chosen in our calculations by taking into account Eq. (\ref{eq:dipole}). We show our results in Table \ref{tab:DtoKdipole}. From Table \ref{tab:DtoKdipole}, one can see that $BR({D^0} \to {\pi ^ - }{\mu ^ + }{\nu _\mu })/BR({D^0} \to {K^ - }{\mu ^ + }{\nu _\mu })$ = 0.072 and $BR({D^0} \to {\pi ^ - }{e^ + }{\nu _e})/BR({D^0} \to {K^ - }{e^ + }{\nu _e})$ = 0.071, which are in agreement with the results of PDG \cite{PDG:2020} with the deviations about 0.033 and 0.167, respectively. Compared the results in Tables \ref{tab:BRDtoK} and \ref{tab:DtoKdipole}, taking Eq. (\ref{IWF:wave function}), which is considered the IWFs in the form factors $F_0(q^2)$ and $F_1(q^2)$, leads to better results in the semileptonic decay widths of ${D^0} \to {\pi^ - }$, as shown in Table \ref{tab:BRDtoK}. In this case, the differences between the third model and experimental values are 0.129 for the electron channel and 0.109 for the muon channel. Thus, one can get the motivating results with the pole dominated form factors for $F_0(q^2)$ and $F_1(q^2)$, see Eq. (\ref{eq:dipole}). For instance, the differences between our results and the experimental values for ${D^0} \to {\pi^ - }{e^ + }{\nu _e}$ are 0.038 and 0.016 for the muon channel. Considering the IWF of Eq. (\ref{IWF:exp}) in terms of the exponential function versus two parameters $\rho^2$ and $\omega$, the obtained results will be closer to experimental measurements for the semileptonic decay widths of ${D^0} \to {K^ - }$, where the differences are 0.298 for the electron channel and 0.258 for the muon channel, whereas in Table \ref{tab:DtoKdipole}, we obtain these differences as 0.031 and 0.091 for the electron and muon channels, consecutively. 

\begin{center}

\begin{table}[ht]
\caption{Branching fractions for ${D^0} \to {K^ - }{l^ + }\nu $  and ${D^0} \to {\pi ^ - }{l^ + }\nu $  (in \%) using the dipole form factor.}
\label{tab:DtoKdipole}
\begin{tabular}{|p{2.5cm}|p{2.5cm}|p{2.5cm}|p{2.5cm}|}
\hline    

\textbf{Decay}  &  \textbf{This work }  & \textbf{Others} & \textbf{Exp. \cite{PDG:2020}}    \\  \hline
${D^0} \to {K^ - }{e^ + }\nu $ & {3.572}  &   3.49, 4.78 \cite{Hwang:1998ph}; 3.2 \cite{Wu:2006rd}  & {$3.541 \pm 0.034$}   \\  \hline
${D^0} \to {K^ - }{\mu ^ + }\nu  $ & {3.501}  &   3.38, 4.67 \cite{Hwang:1998ph}; 3.15 \cite{Wu:2006rd}   & {$ 3.41 \pm 0.04$}  \\  \hline
 ${D^0} \to {\pi ^ - }{e^ + }\nu $ & {0.253}  &   0.292, 0.594 \cite{Hwang:1998ph}; 0.278 \cite{Wu:2006rd}  & {0.291}  \\  \hline
 ${D^0} \to {\pi ^ - }{\mu ^ + }\nu $ & {0.251}  &   0.288, 0.586 \cite{Hwang:1998ph}; 0.275 \cite{Wu:2006rd}  & {0.267}  \\  \hline
\end{tabular} 
\end{table}
\end{center}

\vspace{-1cm}
\section{Conclusions}
\label{section:4}
In the present work, we present a phenomenological potential model based on the harmonic and Yukawa terms to obtain the masses of charmed mesons. We employ the harmonic oscillator wave functions in a variational method, and calculate the decay widths of the pseudoscalar charmed mesons $D$ and $D_s$ decaying into a lepton pair using the Isgur-Wise functions and dipole form factors. We show our results in Tables \ref{tab:massD} to \ref{tab:DtoKdipole} and Figs. \ref{fig:potential} to \ref{subfig2:fourfigs}, and find how the results are influenced by taking different form factors. The form factors play an essential role in evaluating the observables such as the branching fractions, because they are included in the dynamical information of the decay. Taking different models for the form factors of the semileptonic decays of $D$ and $D_s$, lead to different results of the branching fractions. According to the results in Tables \ref{tab:massD}, \ref{tab:massDs}, \ref{tab:masseta}, the mass differences are 9.381 MeV ($D$), 10.253 MeV ($D_s$), 47.452 MeV ($\eta_c(1S)$), 24.848 MeV ($D^*$), 2.738 MeV ($D_s^*$), 13.366 MeV ($D_3^*(2750)$), 3.603 MeV ($D_{s3}^*(2860)^{\pm}$) and 19.342 MeV ($\psi(3842)$) in comparison with the experimental measurements. Our slope parameters of the Isgur-Wise function for the $D$ and $D_s$ mesons are kept in the range of $0.35 \leqslant {\rho ^2} \leqslant 1.15$ \cite{Blok:1992fc}, since we obtain $\rho _D^2 = 0.69$ and $\rho _{{D_s}}^2 = 0.86$. In our present work, using of the third model of the Isgur-Wise function, Eq. (\ref{IWF:wave function}), we calculate the ratios of the semileptonic decay width of $D$ as $BR({D^0} \to {\pi ^ - }{e^ + }{\nu _e})/BR({D^0} \to {K^ - }{e^ + }{\nu _e})$ = 0.041 and  $BR({D^0} \to {\pi ^ - }{\mu ^ + }{\nu _\mu })/BR({D^0} \to {K^ - }{\mu ^ + }{\nu _\mu })$ = 0.041, which are close to the experimental values 0.085 and 0.074, respectively \cite{PDG:2020}. Regarding to the obtained values in Table \ref{tab:D_stoK}, we get $BR(D_s^{} \to {K^0}{e^ + }{\nu _e})$ = ${\text{3}}{\text{.203}} \times {10^{ - 3}}$, which is close to the experimental one, $(3.4 \pm 0.4) \times {10^{ - 3}}$ \cite{PDG:2020}. We also obtain $BR(D_s^{} \to \eta {\mu ^ + }{\nu _\mu })$ = 2.905 \% compared with the $({\text{2}}{\text{.4}} \pm {\text{0}}{\text{.5}})\%$ one reported in Ref. \cite{PDG:2020}. Using Eq. (\ref{eq:dipole}) for the form factors of the semileptonic $D$ mesons, we calculate $BR({D^0} \to {K^ - }{e ^ + }\nu_ e  )$ = 3.572 $\%$ and $BR({D^0} \to {\pi ^ - }{\mu ^ + }{\nu _\mu })/BR({D^0} \to {K^ - }{\mu ^ + }{\nu _\mu })$ = = 0.072, which are in agreement with experimental values, 3.541 \% and 0.074, respectively. Thus, our results are meaningful for the experimental measurements.

\end{document}